\documentclass{article}
\usepackage{spconf,amsmath,graphicx,hyperref}
\usepackage{subcaption} 
\usepackage{graphicx} 
\usepackage{booktabs}
\usepackage{multirow} 
\usepackage[font=small,labelfont=bf]{caption}

\title{Exploring How Audio Effects Alter Emotion with Foundation Models}
\name{Stelios Katsis, Vassilis Lyberatos, Spyridon Kantarelis, Edmund Dervakos, and Giorgos Stamou}
\address{National Technical University of Athens, Athens, Greece \\
\texttt{stelioskatsis12@gmail.com,\{vaslyb, spyroskanta\}@ails.ece.ntua.gr,} \\ \texttt{eddiedervakos@islab.ntua.gr, gstam@cs.ntua.gr}}

\begin{document}
\small

%
\maketitle
\begin{abstract}
Audio effects (FX) such as reverberation, distortion, modulation, and dynamic range processing play a pivotal role in shaping emotional responses during music listening. While prior studies have examined links between low-level audio features and affective perception, the systematic impact of audio FX on emotion remains underexplored. This work investigates how foundation models—large-scale neural architectures pretrained on multimodal data—can be leveraged to analyze these effects. Such models encode rich associations between musical structure, timbre, and affective meaning, offering a powerful framework for probing the emotional consequences of sound design techniques. By applying various probing methods to embeddings from deep learning models, we examine the complex, nonlinear relationships between audio FX and estimated emotion, uncovering patterns tied to specific effects and evaluating the robustness of foundation audio models. Our findings aim to advance understanding of the perceptual impact of audio production practices, with implications for music cognition, performance, and affective computing.
\end{abstract}
\section{Introduction}

The relationship between sound and human emotion has been extensively studied across multiple domains, particularly within affective computing and music cognition. Prior research has demonstrated how low-level audio features—such as timbre, tempo, pitch, and rhythm—correlate with emotional responses in listeners \cite{laurier2009exploring, panda2020audio, lyberatos2024perceptual, 10944805,lyberatos2025musicinterpretationemotionperception}. Parallel investigations have examined how environmental auditory conditions, including soundscape complexity and the number of simultaneous sound sources, influence affective perception \cite{han2023effect}. However, there remains a notable gap in understanding how audio FX—such as reverberation, distortion, modulation, and dynamic range processing—systematically alter emotional responses in music listening contexts.

Unlike fundamental audio features, audio FX are intentional sound design tools commonly applied in music production and performance to modulate aesthetic and emotional impact. Yet, their role in shaping affective perception has been underexplored. For instance, \cite{wilkie2025reverberation} showed that longer reverberation times in recorded music increase ratings of “sublimity,” including emotional qualities like awe and nostalgia. Similarly, distortion has been linked to heightened arousal and affective intensity, potentially evoking excitement or even aggression \cite{panda2020audio}. Higher rates of vibrato are linked to elevated arousal levels \cite{scherer2015comparing, song2024emotional}. Additionally, an increase in the number of perceived sound sources correlates with increased arousal but may reduce pleasantness \cite{han2023effect}.

Despite these isolated findings, there is no comprehensive study that systematically investigates how diverse audio FX in music—individually or in combination—impact emotional perception. Given that audio FX are central to contemporary music production, a focused analysis of their affective consequences is both timely and necessary. 

At the same time, advances in foundation models—large-scale neural architectures pre-trained on multimodal datasets, have demonstrated an ability to capture high-level semantic and perceptual relationships across domains, including with audio FX \cite{deng2025investigating,shah2021all}. In the music domain, foundation models such as MusicLM \cite{agostinelli2023musiclm}, MERT \cite{li2023mert}, Qwen \cite{qwen2023}, and CLAP \cite{elizalde2022clap} have shown promising abilities in generating emotionally expressive audio, learning disentangled latent representations, and enabling cross-modal retrieval tasks~\cite{akman2025audio}. These models implicitly encode rich associations between musical structure, timbre, and affective meaning, making them ideal tools to probe novel hypotheses about music-emotion relationships.

In this work, we leverage foundation models to investigate how specific audio FX influence the way foundation models perceive emotions in music. We analyze performance changes, shifts in predicted emotional labels and dimensions, and alterations in model embeddings across three state-of-the-art foundation models and diverse datasets with both categorical and dimensional annotations, using six common audio FX at different intensity levels, including fixed real-world. By combining controlled manipulations with these realistic production scenarios, we identify correlations between audio FX and estimated emotions and uncover properties of how foundation models represent these affective cues.
\begin{table*}[h!]
\centering
\tiny
\captionsetup{justification=centering}
\caption{Audio FX $\Delta$ (Level$_{10}$-Level$_{1}$) across MSE Arousal, MSE Valence, and F1 score. Bold values indicate the largest absolute difference among audio FXs for a specific model and dataset, while underlined values indicate the largest absolute difference among models for a specific audio FX and dataset.}
\renewcommand{\arraystretch}{1.3}\setlength{\tabcolsep}{6pt}
\begin{tabular}{l|ccc|ccc|ccc|ccc|ccc|ccc}
\toprule
\multirow{3}{*}{\textbf{Audio FX}} &
\multicolumn{6}{c|}{\textbf{MSE Arousal}} &
\multicolumn{6}{c|}{\textbf{MSE Valence}} &
\multicolumn{6}{c}{\textbf{F1 Score Categorical}} \\
\cmidrule(lr){2-7}\cmidrule(lr){8-13}\cmidrule(lr){14-19}
& \multicolumn{3}{c|}{\textbf{DEAM}} & \multicolumn{3}{c|}{\textbf{witheFlow}}
& \multicolumn{3}{c|}{\textbf{DEAM}} & \multicolumn{3}{c|}{\textbf{witheFlow}}
& \multicolumn{3}{c|}{\textbf{witheFlow}} & \multicolumn{3}{c}{\textbf{EMOPIA}} \\
\cmidrule(lr){2-4}\cmidrule(lr){5-7}\cmidrule(lr){8-10}\cmidrule(lr){11-13}\cmidrule(lr){14-16}\cmidrule(lr){17-19}
& \textbf{Qwen} & \textbf{MERT} & \textbf{CLAP}
& \textbf{Qwen} & \textbf{MERT} & \textbf{CLAP}
& \textbf{Qwen} & \textbf{MERT} & \textbf{CLAP}
& \textbf{Qwen} & \textbf{MERT} & \textbf{CLAP}
& \textbf{Qwen} & \textbf{MERT} & \textbf{CLAP}
& \textbf{Qwen} & \textbf{MERT} & \textbf{CLAP} \\
\midrule
Chorus     & +0.003 & \underline{+0.008} & -0.001 & +0.060 & \underline{+0.110} & +0.029 
           & \underline{+0.007} & +0.004 & -0.001 & \underline{+0.120} & \underline{+0.120} & +0.070
           & -0.160 & \underline{-0.190} & -0.090 & \underline{-0.310} & -0.110 & -0.070 \\
Distortion & +0.012 & +0.023 & \underline{\textbf{+0.028}} & \underline{+0.160} & +0.122 & -0.027
           & +0.009 & \textbf{+0.017} & \textbf{\underline{+0.027}} & \textbf{+0.240} & \textbf{\underline{+0.210}} & \textbf{+0.110}
           & \underline{\textbf{-0.488}} & \textbf{-0.433} & \textbf{-0.149} & \textbf{-0.370} & -\underline{\textbf{0.390}} & \textbf{-0.350} \\
Phaser     & \underline{+0.006} & +0.005 & +0.002 & \underline{\textbf{+0.171}} & +0.129 & -0.040
           & \underline{+0.006} & \underline{+0.006} & +0.001 & +0.075 & \underline{+0.168} & +0.030
           & \underline{-0.266} & -0.230 & -0.041 & \underline{-0.180} & -0.100 & -0.020 \\
EQ         & +0.007 & +0.010 & \underline{+0.014} & +0.051 & \underline{+0.060} & -0.033
           & +0.005 & +0.007 & \underline{+0.011} & +0.050 & \underline{+0.140} & -0.050
           & -0.103 & \underline{-0.186} & +0.041 & +0.040 & \underline{-0.060} & +0.030 \\
Delay      & +0.003 & \underline{+0.004} & -0.001 & +0.016 & \underline{+0.113} & +0.008
           & +0.005 & \underline{+0.005} & -0.002 & \underline{+0.060} & \underline{+0.060} & +0.040
           & -0.091 & \underline{-0.147} & +0.000 & \underline{-0.200} & -0.090 & +0.100 \\
Reverb     & \underline{\textbf{+0.021}} & \textbf{+0.012} & +0.010 & -0.012 & \underline{\textbf{+0.141}} & \textbf{+0.106}
           & \textbf{\underline{+0.020}} & +0.015 & +0.006 & +0.030 & \underline{+0.060} & +0.055
           & -0.022 & \underline{-0.150} & -0.116 & -0.160 & \underline{-0.230} & -0.170 \\
\bottomrule
\end{tabular}
\label{tab:audiofx_all}
\end{table*}

\section{Material}\label{sec:mat}

In this study, we employed a diverse set of datasets and models, spanning both deep and shallow architectures, along with several commonly used audio FXs with varying parameter settings. The code, analytical details, and complete experimental results are available in our GitHub repository~\footnote{https://github.com/stelioskt/audioFX}.

We utilized three \textbf{datasets} capturing both categorical and dimensional emotional annotations: \textit{EMOPIA}~\cite{hung2021emopia}, \textit{DEAM}~\cite{alajanki2016benchmarking}, and \textit{witheFlow}~\cite{lyberatos2025musicinterpretationemotionperception}. The \textit{EMOPIA} dataset is a multimodal (audio and MIDI) resource focused on perceived emotion in pop piano music, containing 1,087 clips from 387 songs. Each clip is annotated with one of four categorical emotions—\textit{Excitement, Anger, Sadness}, or \textit{Calmness}—supporting both clip-level classification and song-level emotion analysis. The \textit{DEAM} dataset comprises 1,802 excerpts and full songs annotated with valence and arousal as continuous, time-varying values per second, as well as aggregated ratings, along with metadata on duration, genre, and folksonomy tags, enabling regression-based modeling of emotion dynamics. The part of the \textit{witheFlow} dataset we used includes 235 solo-instrument recordings, annotated by 20 listeners with continuous valence-arousal values per second following the \textit{Circumplex Model of Affect}~\cite{russell1980circumplex}, as well as multi-label categorical tags based on the \textit{Geneva Emotional Music Scale (GEMS-9)}~\cite{jacobsen2024assessing}, making it suitable for both dimensional and categorical emotion modeling in music.

\begin{figure}[!b]
    \centering

    \begin{subfigure}[b]{0.158\textwidth}
        \centering
        \includegraphics[width=\linewidth]{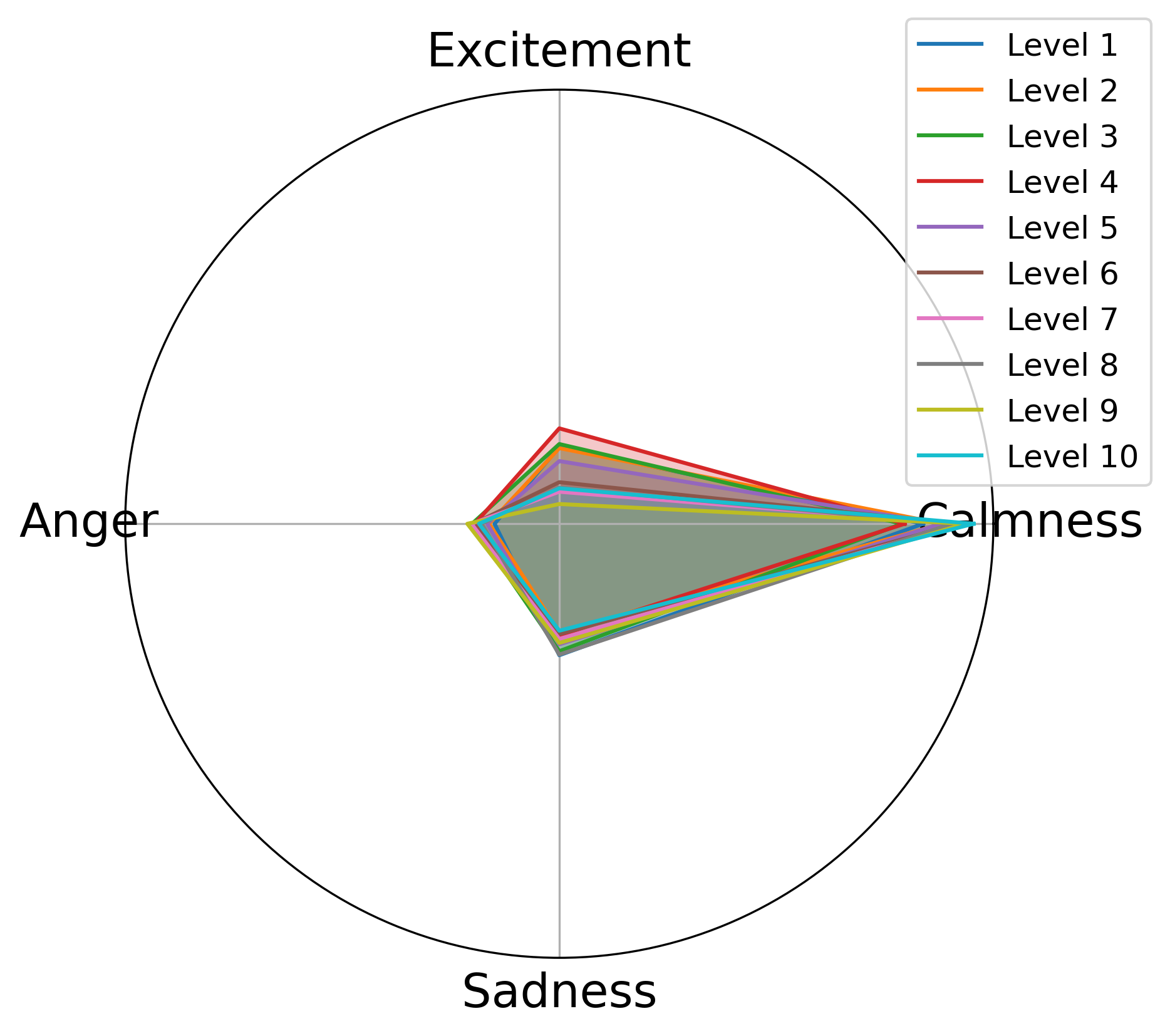}
        \caption{CLAP-Chorus}
    \end{subfigure}
    \begin{subfigure}[b]{0.158\textwidth}
        \centering
        \includegraphics[width=\linewidth]{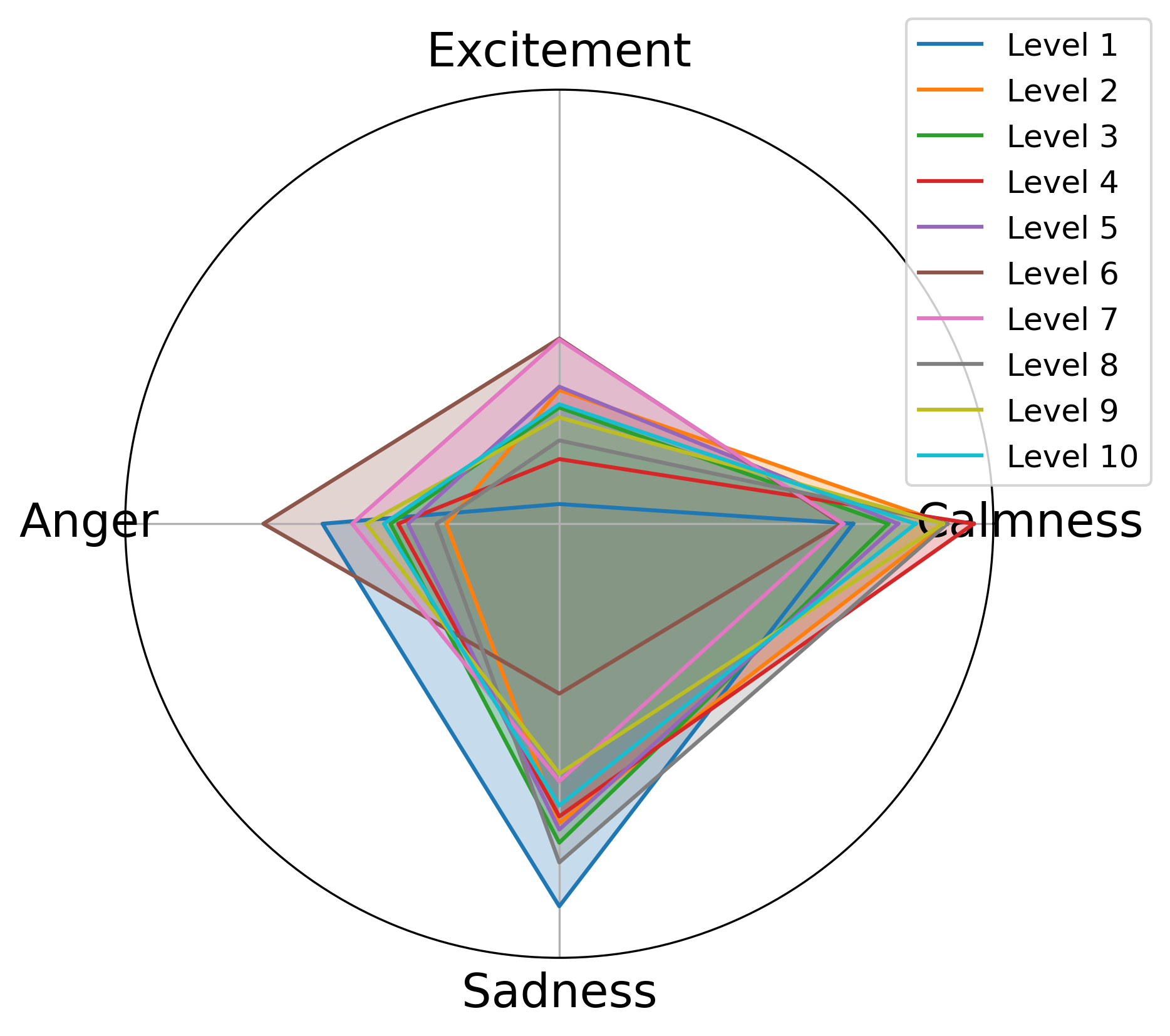}
        \caption{CLAP-Delay}
    \end{subfigure}
    \begin{subfigure}[b]{0.158\textwidth}
        \centering
        \includegraphics[width=\linewidth]{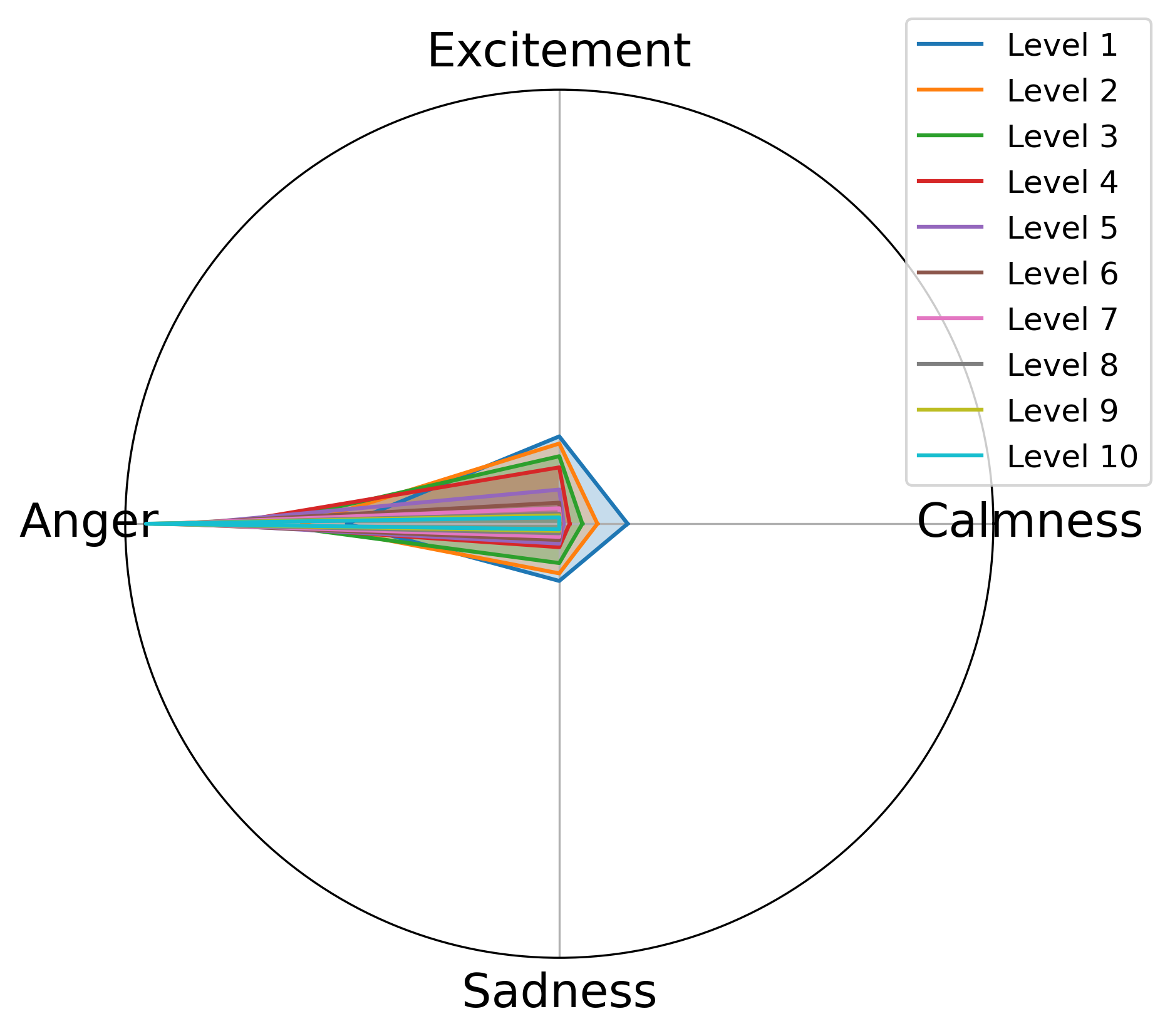}
        \caption{CLAP-Distortion}
    \end{subfigure}

    \vspace{0.8em} 

    \begin{subfigure}[b]{0.158\textwidth}
        \centering
        \includegraphics[width=\linewidth]{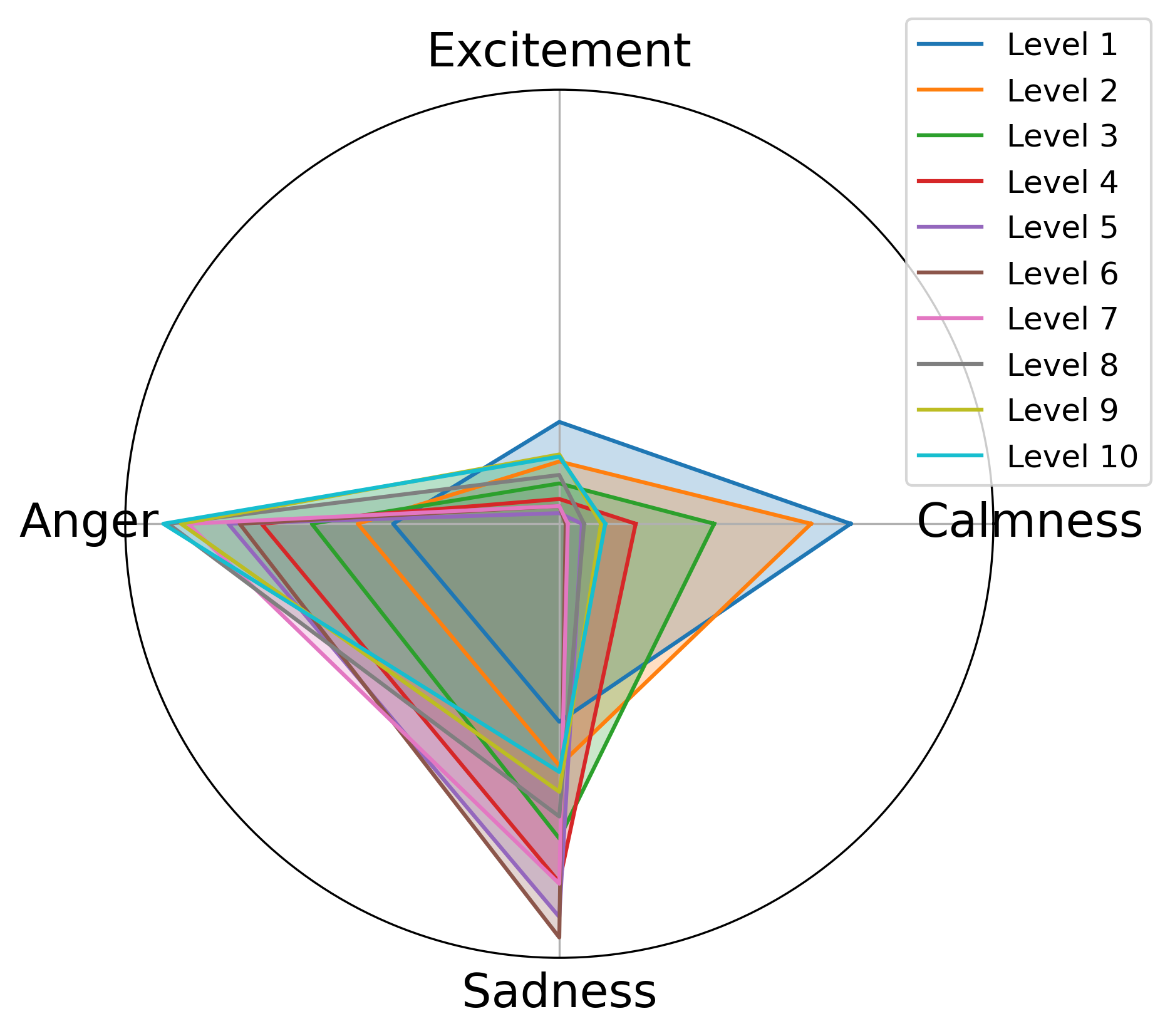}
        \caption{Qwen-Chorus}
    \end{subfigure}
    \begin{subfigure}[b]{0.158\textwidth}
        \centering
        \includegraphics[width=\linewidth]{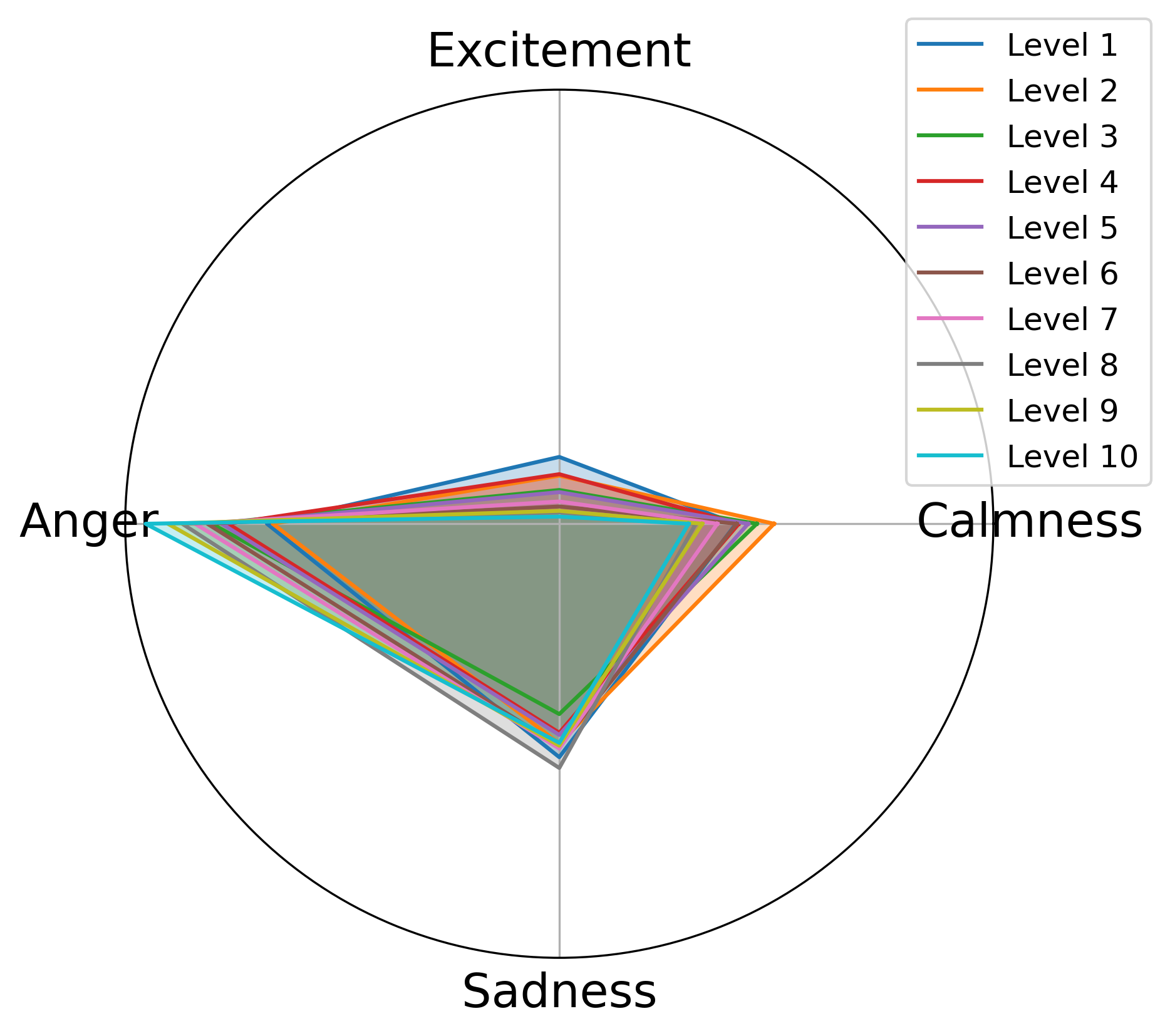}
        \caption{Qwen-Delay}
    \end{subfigure}
    \begin{subfigure}[b]{0.158\textwidth}
        \centering
        \includegraphics[width=\linewidth]{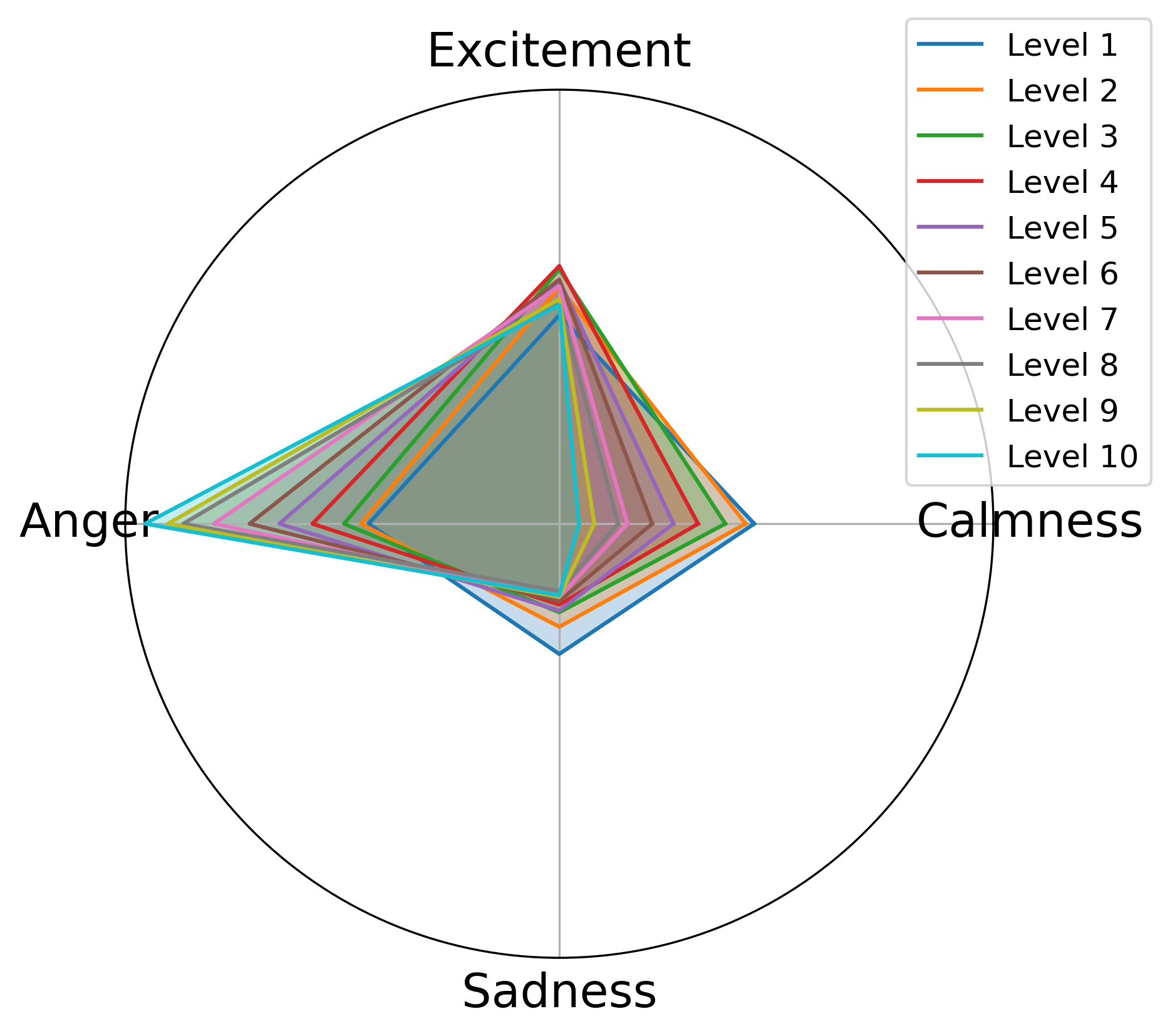}
        \caption{Qwen-Distortion}
    \end{subfigure}

    \vspace{0.8em}

    \begin{subfigure}[b]{0.158\textwidth}
        \centering
        \includegraphics[width=\linewidth]{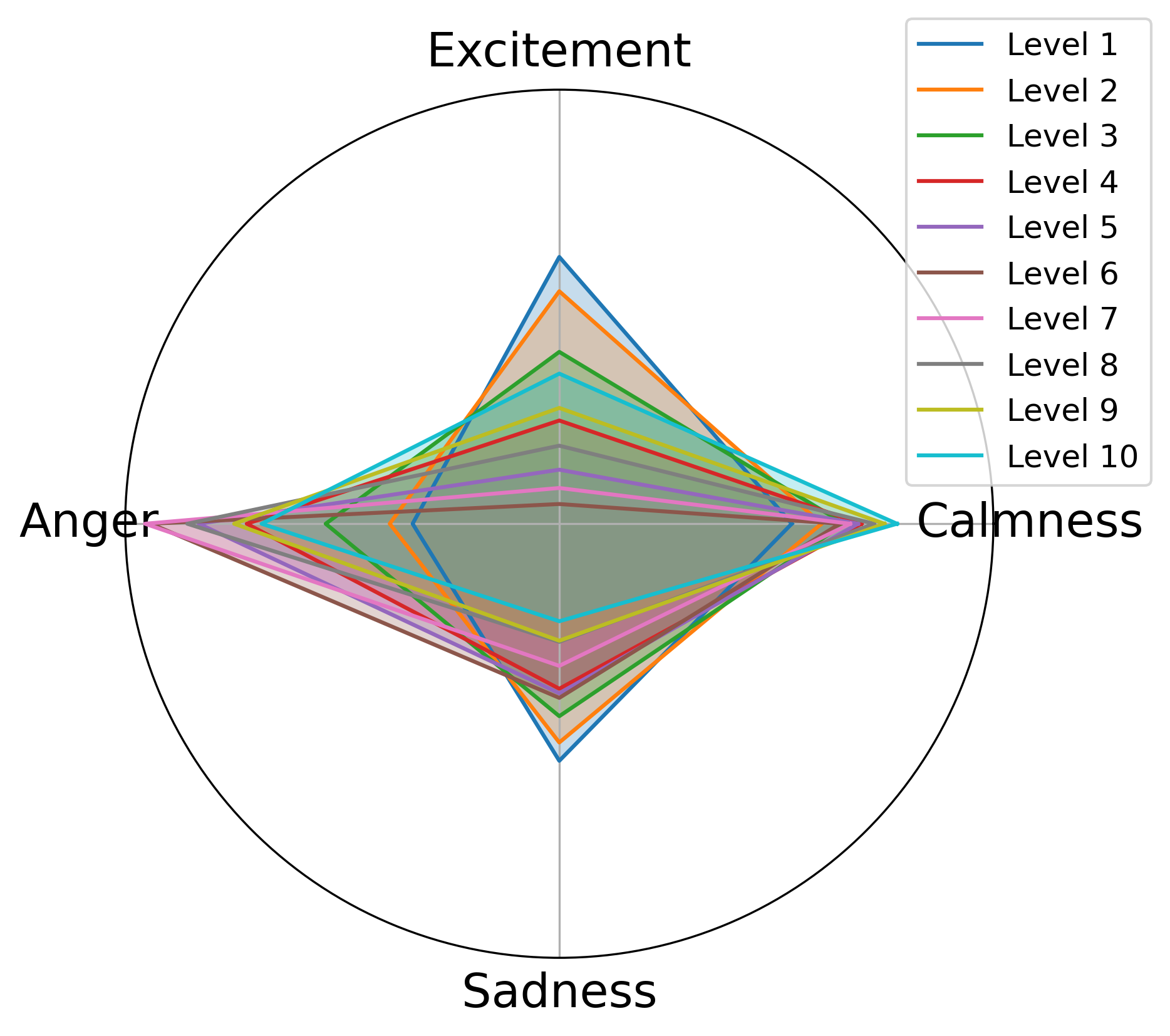}
        \caption{MERT-Chorus}
    \end{subfigure}
    \begin{subfigure}[b]{0.158\textwidth}
        \centering
        \includegraphics[width=\linewidth]{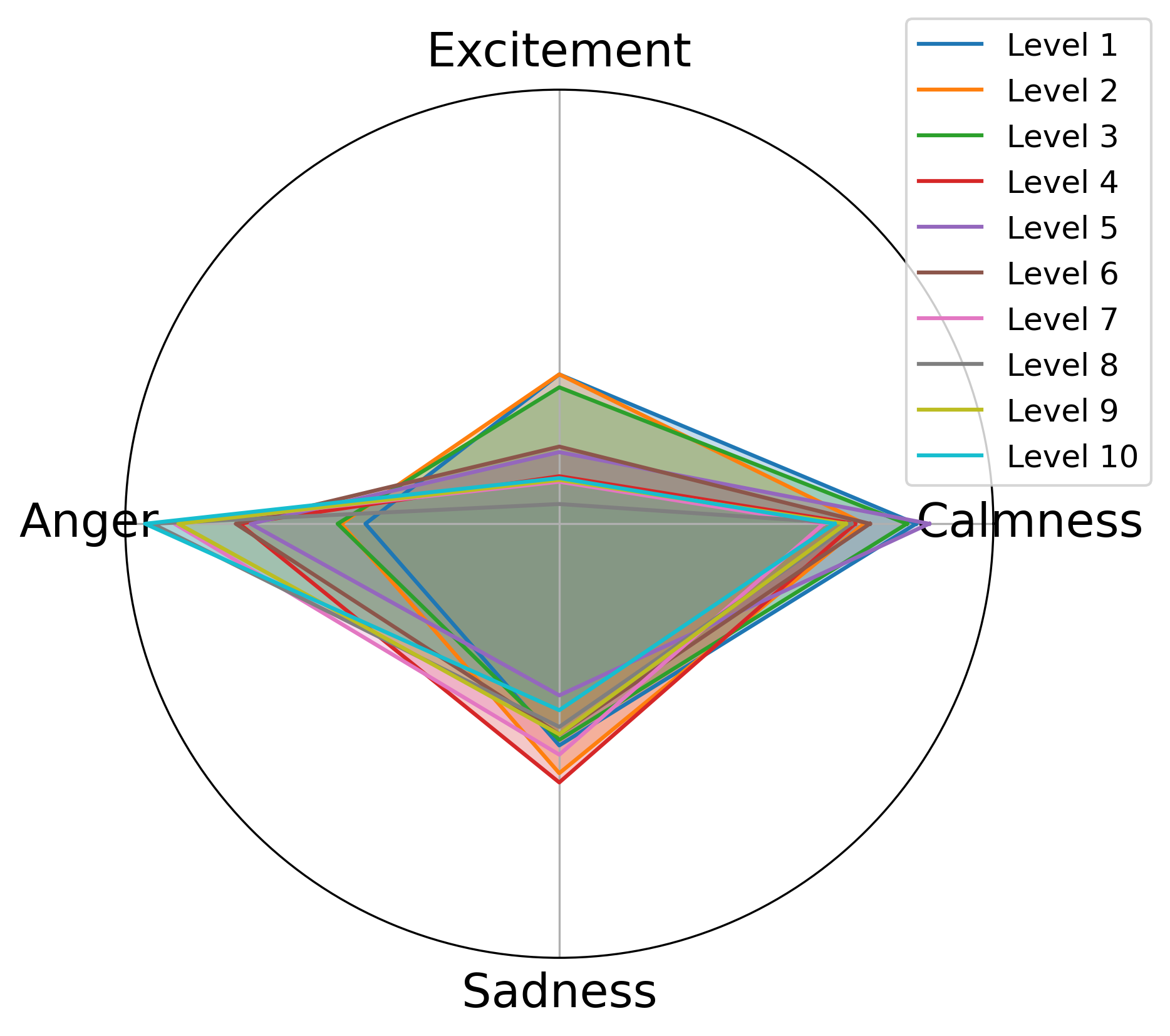}
        \caption{MERT-Delay}
    \end{subfigure}
    \begin{subfigure}[b]{0.158\textwidth}
        \centering
        \includegraphics[width=\linewidth]{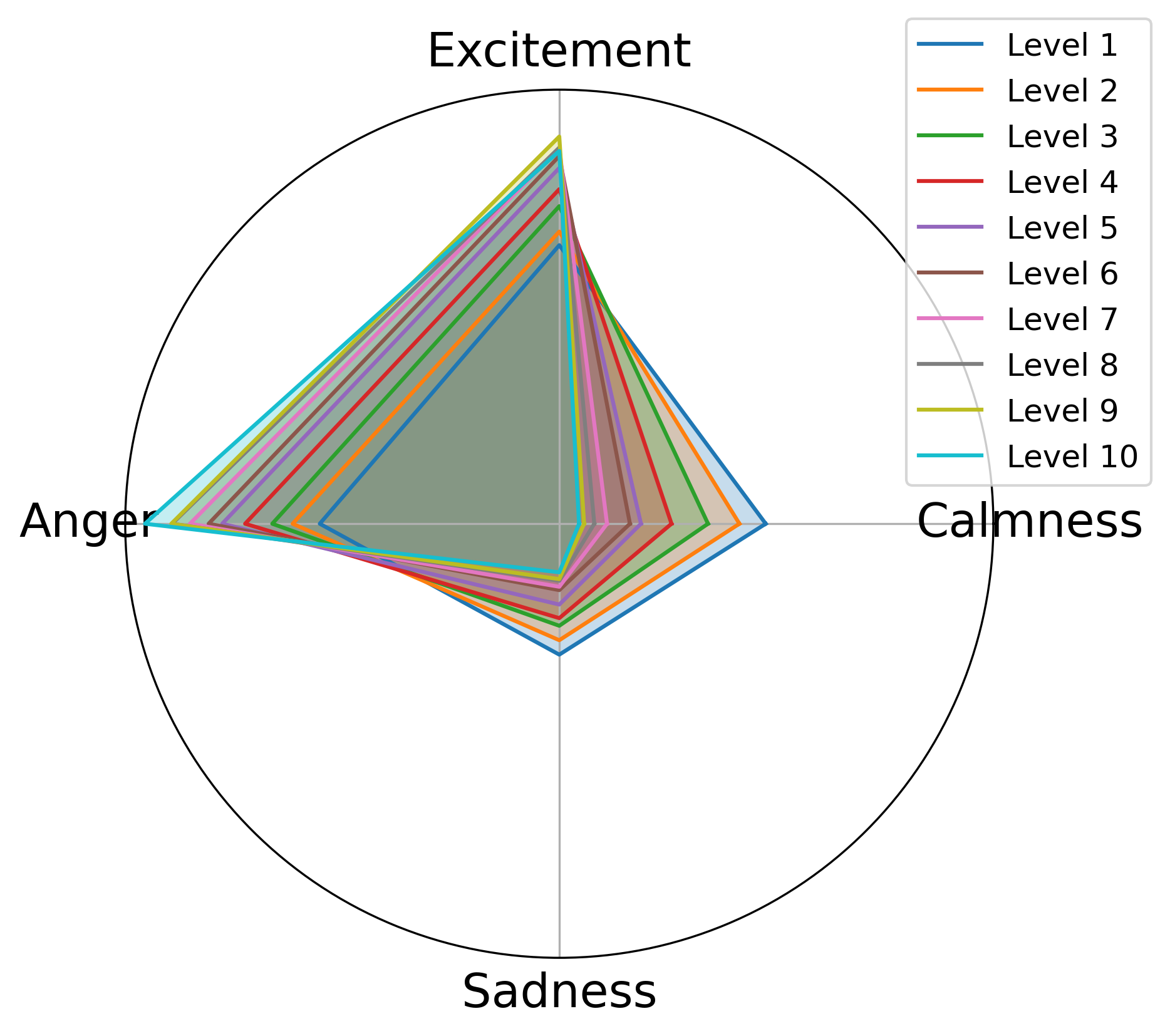}
        \caption{MERT-Distortion}
    \end{subfigure}

    \caption{Radar plots of emotion predictions for CLAP, Qwen, and MERT across three audio effects for the \textit{EMOPIA} dataset. Each level in the plots depicts the distribution of emotions, normalised based on the greatest value in a given plot.}
    \label{fig:radar_grid}
\end{figure}

To investigate and interpret musical emotion, we employed three state-of-the-art \textbf{foundation models}, each offering unique capabilities. First, \textit{MERT-v1-330M}~\cite{li2023mert} is a 330M-parameter model (24 layers with 1024 hidden units) trained on 160,000 hours of music audio using masked language modeling with RVQ-VAE and CQT teacher signals. MERT operates on 24 kHz audio at 75 frames per second and is particularly suited to capturing fine-grained musical and acoustic emotion cues. Second, \textit{CLAP}~\cite{wu2023large} is a dual-encoder model with roughly 630M parameters, combining a Transformer-based (HTS-AT) audio encoder with a \textit{RoBERTa}-based text encoder. CLAP was trained on 128K–630K audio–text pairs via contrastive learning, using log-Mel spectrogram inputs at 48 kHz to learn a shared audio–text representation space with strong zero-shot generalization. Finally, \textit{Qwen2-Audio-7B}~\cite{chu2024qwen2} is a 7B-parameter model featuring a \textit{Whisper}-style audio encoder and \textit{Qwen}-style decoder. Qwen2-Audio-7B was trained on over 300K hours of speech, music, and environmental audio to support broad audio-language tasks, including automatic speech recognition, classification, question answering, and conversational interaction.

For the head classifiers of our foundation models, we employed shallow \textbf{interpretable models} tailored to each task and dataset, training them while keeping the foundation model frozen. For regression on \textit{DEAM} and \textit{witheFlow} valence-arousal annotations, we used an \textit{XGBRegressor}~\cite{chen2016xgboost} to predict continuous emotional dimensions, evaluating performance with MAE, MSE, and $R^2$. For single-label classification on \textit{EMOPIA}, we applied an \textit{XGBClassifier}, assessing weighted accuracy, precision, recall, and F1-score. For multi-label classification on \textit{witheFlow} categorical tags, we implemented a \textit{OneVsRest} strategy with \textit{XGBClassifier}, measuring F1-micro and F1-macro scores with a threshold of 0.5. This configuration allowed us to effectively adapt gradient-boosted tree models to both dimensional and categorical emotion prediction tasks.

\begin{figure*}[!t]
    \centering
    \includegraphics[width=1\textwidth]{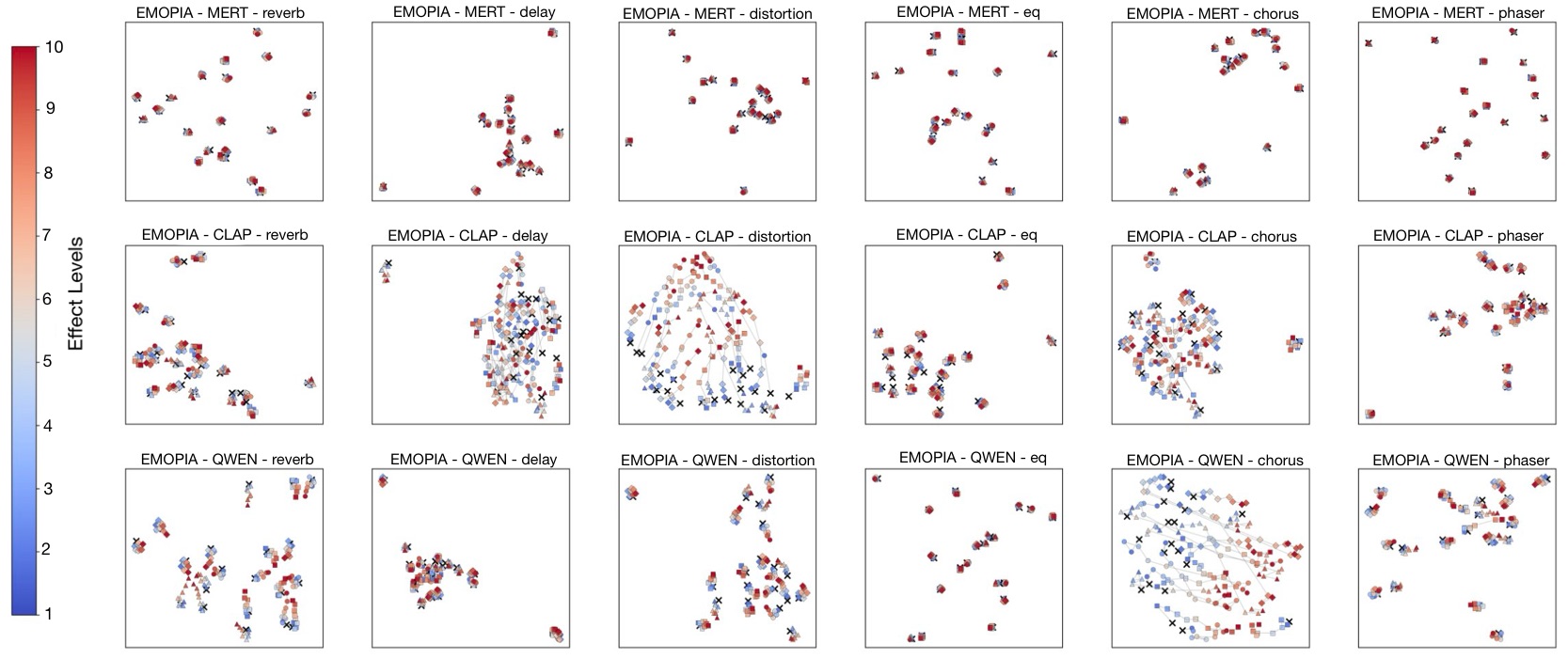}
    \caption{UMAP visualization of relative distance in the most emotion-discriminative foundation model embeddings for the \textit{EMOPIA} dataset, showing trajectories generated for each input identity after applying audio FX with the intensity ranging from 1 to 10.}
    \label{fig:emopia_trajectories}
\end{figure*}

To apply controlled alterations to our input and observe the behavior of our chosen foundation model, we used six \textbf{audio FX} from the pedalboard library, each scaled in intensity from 1 to 10. Reverb simulates acoustic space by adjusting \textit{room\_size}, creating the perception of different environments from small rooms to large halls, which can influence temporal and spectral characteristics of the signal. Delay introduces echoes through \textit{delay\_seconds} and \textit{feedback}, affecting rhythmic perception and reinforcing certain frequencies. Distortion adds harmonic saturation via \textit{drive\_db}, enhancing harmonic content and introducing nonlinearities that challenge the model’s robustness. EQ shapes the frequency spectrum with \textit{low\_cutoff} and \textit{high\_cutoff} filters, emphasizing or attenuating spectral bands and altering timbral balance. Chorus thickens the sound by duplicating and slightly detuning the signal with \textit{rate}, \textit{depth}, and \textit{feedback}, producing subtle modulation that can affect perceived pitch and texture. Phaser applies sweeping frequency modulation via \textit{rate}, \textit{depth}, and \textit{feedback}, generating moving notches that alter harmonic relationships over time.

\section{Experiments}
Using the above material, we conducted four experiments to investigate the relationships between audio effects and estimated emotions. First, we analyzed performance changes to assess how audio manipulations impact model accuracy. Second, we examined shifts in emotion predictions to identify changes in predicted emotional labels or dimensions. Third, we explored embedding space alterations to understand how audio effects influence the internal representations learned by the models. Finally, we investigated how real-world audio FX settings affect emotion estimation.

\subsection{Impact on Model Performance}

Inspired by \cite{deng2025investigating}, that studies the impact of audio effects on foundation models, and \cite{albin2015objective}, that examines performance changes in speech emotion recognition models induced by codec processing, we aimed to explore how incorporating emotional information affects performance in emotion recognition tasks. To investigate this, we trained interpretable machine learning classifiers using embeddings extracted from foundation models across all datasets. This approach allowed us to develop models for multi-class classification, multi-label classification, and regression tasks.

After training our interpretable classifiers, we applied various audio effects at different intensity levels and analyzed their impact on model performance. All foundation models were consistently used as backbones across tasks and datasets. We hypothesized that any observed performance degradation would indicate a potential correlation between the audio FX and the estimated emotion.

Overall, we observed a general decrease in performance, indicating that audio effects may influence emotion recognition. As summarized in Table~\ref{tab:audiofx_all}, phaser and distortion in particular evoked the largest performance declines across multiple intensity levels. All three models exhibited similar trends, with the decrease becoming more pronounced as the intensity of the audio effects increased.

\subsection{Shifts in Emotion Predictions}

Previously trained pipelines were used to run inference on audio samples that had been modified with six common effects applied at varying intensities. Transparent classifiers built on different foundation model backbones generated emotion predictions from these altered inputs (see Section~\ref{sec:mat}). The resulting predictions were compared with those from the original samples, and the observed changes were visualized to reveal how the effects shifted predicted emotions. This approach enabled the examination of potential correlations between specific audio manipulations and estimated emotional responses.

In Fig.~\ref{fig:radar_grid}, we examine how emotional predictions in the \textit{EMOPIA} dataset are affected by different audio effects. Notably, high levels of distortion consistently increase \textit{Anger} predictions while decreasing \textit{Calmness} across all models~\cite{gao2022black}. In contrast, delay and chorus effects introduce greater variability, suggesting that these manipulations create ambiguity in the models’ emotional interpretations~\cite{cesarini2024reverb}. Interestingly, increasing the chorus effect tends to boost \textit{Calmness} predictions in the CLAP and MERT models, whereas increasing the delay effect raises \textit{Anger} predictions in the CLAP and Qwen models.

\subsection{Embedding Space Alterations}

In this section, we present an experiment investigating how the embeddings of a foundation model—representations that encode emotional information—are altered as common audio effects are gradually applied. Inspired by the visualization approach of Deng et al.~\cite{deng2025investigating}, we identified the most important features from our shallow emotion classification model. Using the top 25 features, we performed UMAP~\cite{mcinnes2018umap} dimensionality reduction and visualized the foundation model's embeddings for inputs modified at different effect intensities. This approach allowed us to observe how audio effects reshape the structure of the embeddings and to identify which models are most sensitive to these changes.

To ensure stability and comparability, embeddings were first standardized per (dataset, model) using zero-mean/unit-variance scaling, then cleaned via (i) variance thresholding (removing features with variance \(< 1 \times 10^{-6}\)) and (ii) iterative correlation pruning (dropping features with \(|r| > 0.95\)). Feature selection was then performed: for regression-style datasets (\textit{DEAM}, \textit{witheFlow}) we used \textit{ElasticNetCV} with \(l1\_ratio \in \{0.5, 0.8\}\), \(\alpha\) values log-spaced from \(10^{-3}\) to \(10^{2}\) (50 values), 5-fold cross-validation, \(\max\_iter = 60000\), and \(tol = 2 \times 10^{-3}\), while for classification-style datasets (\textit{EMOPIA}, \textit{witheFlow}) we used logistic regression with an elastic net penalty (\(l1\_ratio = 0.5\)), the SAGA solver, \(C = 0.5\), \(max\_iter = 6000\), and \(tol = 2 \times 10^{-3}\). For visualization, we applied UMAP using a cosine metric and spectral initialization; in this experiment, these shallow models were chosen for their linearity and interpretability, with \(K = 25\), \(n\_neighbors = 30\), and \(min\_dist = 0.5\). To avoid crowding and label imbalance, we sampled 20 tracks per effect for regression datasets, 5 tracks per label for classification datasets, and 3 tracks per label for the multi-label dataset (witheFlow), using a fixed random seed (\(42\)), covering the effects reverb, delay, distortion, EQ, chorus, and phaser.

In Fig.~\ref{fig:emopia_trajectories}, we show how embeddings from each foundation model respond to audio effects on the \textit{EMOPIA} dataset. Our analysis reveals systematic differences in how foundation models encode the impact of audio effects. CLAP embeddings exhibit large, structured displacements as FX levels increase—particularly for delay, chorus, and distortion—indicating strong sensitivity to timbral manipulations. Qwen also shows noticeable shifts, though with less consistent trajectories, whereas MERT remains relatively stable across all effect levels, suggesting robustness to such manipulations, likely due to its training on music-specific tasks. Overall, trajectory length and variance in embedding space can serve as metrics for quantifying the influence of audio FX on emotion, with the magnitude of this impact depending strongly on the choice of foundation model.

\subsection{Real World Scenario}

We tested real-world effect chains designed to mimic iconic sounds from Pink Floyd, U2 (emphasizing reverb and delay to create atmospheric textures), and Rage Against the Machine (heavy distortion with moderate chorus)~\footnote{https://www.kitrae.net/music/David\_Gilmour\_Effects\_And\_Gear.html}~\footnote{https://www.guitarchalk.com/guitar-amp-settings-guide/\#t-1649265629386}. The goal of this experiment was to examine whether, in real-world scenarios, carefully crafted audio FX chains can produce stronger shifts in emotion predictions due to their intentional artistic design.

\begin{figure}[!b]
    \centering
    \includegraphics[width=0.49\textwidth]{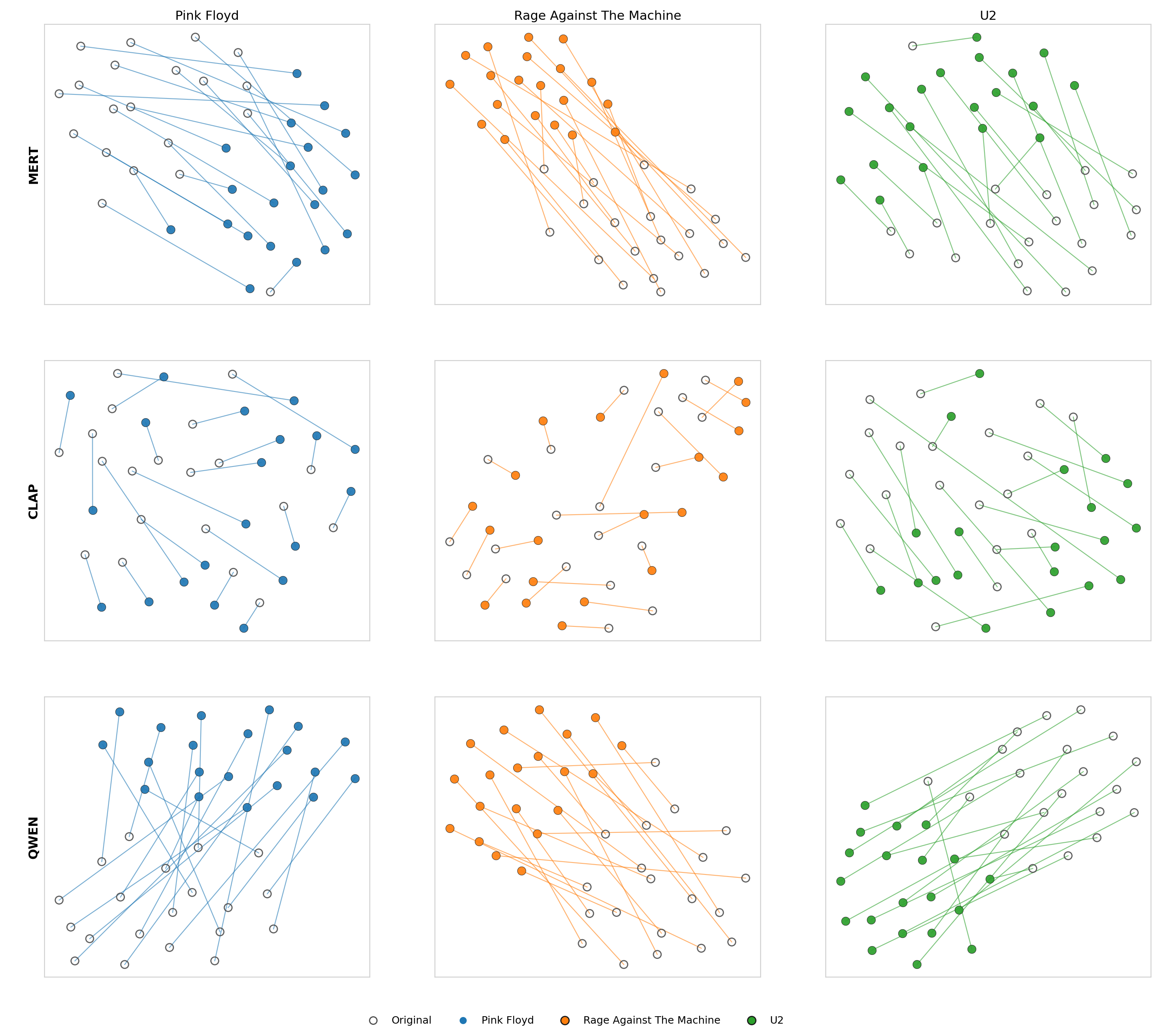}
    \caption{UMAP visualization of relative distance in the most emotion-discriminative foundation model embeddings for the \textit{witheFlow} dataset, showing trajectories generated for each input 
    after applying real-world scenario audio FX.}

    \label{fig:embedding_scenarios}
\end{figure}

As shown in Fig.~\ref{fig:embedding_scenarios}, moving from isolated effects to complete FX chains reveals how cumulative production choices shift embeddings in latent space. MERT and Qwen produce similar trajectories, both showing structured, directional shifts that reflect the cumulative impact of effects. The distortion-heavy chain of Rage Against the Machine yields almost unidirectional trajectories, consistent with strong spectral shaping imposing a uniform transformation. Reverb- and delay-heavy chains of U2 and Pink Floyd also follow a generally consistent pattern, though some divergent paths reflect the more complex ways spatial and temporal effects influence latent representations. CLAP, in contrast, shows shorter, scattered movements, suggesting that its sensitivity is dampened when effects are combined. Overall, real-world effect chains generate larger, more coherent shifts, highlighting that artists naturally select combinations to evoke stronger emotional responses. The nature of these trajectories—unidirectional versus multidirectional—thus serves as a proxy for how different categories of audio effects shape emotional perception in foundation models.

\section{Conclusions and Future Work}

Our study demonstrates that audio effects noticeably alter estimated emotion in music. Distortion, in particular, strongly increase \textit{Anger} while reducing \textit{Calmness}, whereas chorus and delay introduce more variability in predictions. Analysis of embedding-space trajectories shows that the magnitude and structure of latent-space shifts reflect the emotional impact of FX, with real-world effect chains producing the most pronounced and coherent changes, likely because they are intentionally designed by artists to evoke strong emotional responses. Among the models, MERT exhibited relative robustness, while CLAP and Qwen were more sensitive to manipulations.  

For future work, we plan to investigate the simultaneous application of multiple audio effects, extend the study to additional datasets and foundation models, explore alternative probing techniques to gain deeper insight into how complex audio transformations influence emotional perception in music, and include comparisons to classical feature-based systems as well as analyses of human ground truth after effects.
\clearpage

\section{Acknowledgments}

This work was supported by AWS resources provided by the National Infrastructures for Research and Technology (GRNET), funded by the EU Recovery and Resilience Facility. The research project was implemented within the framework of the H.F.R.I. call “Basic Research Financing (Horizontal Support of All Sciences)” under the National Recovery and Resilience Plan “Greece 2.0,” funded by the European Union – NextGenerationEU (H.F.R.I. Project No. 15111: Emotional Artificial Intelligence in Music Expression).

\bibliographystyle{IEEEbib}
{\footnotesize \bibliography{refs}}

@misc{lyberatos2025musicinterpretationemotionperception,
      title={Music Interpretation and Emotion Perception: A Computational and Neurophysiological Investigation}, 
      author={Vassilis Lyberatos and Spyridon Kantarelis and Ioanna Zioga and Christina Anagnostopoulou and Giorgos Stamou and Anastasia Georgaki},
      year={2025},
      eprint={2506.01982},
      archivePrefix={arXiv},
      primaryClass={cs.HC},
      url={https://arxiv.org/abs/2506.01982}, 
}

@article{wilkie2025reverberation,
  title={Reverberation Time and Musical Emotion in Recorded Music Listening},
  author={Wilkie, Hannah and Harrison, Peter},
  journal={Music Perception: An Interdisciplinary Journal},
  volume={42},
  number={4},
  pages={329--344},
  year={2025},
  publisher={University of California Press}
}

@article{scherer2015comparing,
  title={Comparing the acoustic expression of emotion in the speaking and the singing voice},
  author={Scherer, Klaus R and Sundberg, Johan and Tamarit, Lucas and Salom{\~a}o, Gl{\'a}ucia L},
  journal={Computer Speech \& Language},
  volume={29},
  number={1},
  pages={218--235},
  year={2015},
  publisher={Elsevier}
}

@inproceedings{song2024emotional,
  title={The emotional characteristics of the violin with different pitches, dynamics, and vibrato},
  author={Song, Wenyi and Dinh, Anh-Dung and Horner, Andrew Brian},
  booktitle={Proceedings of Meetings on Acoustics},
  volume={55},
  number={1},
  pages={035004},
  year={2024},
  organization={Acoustical Society of America}
}

@ARTICLE{10944805,
  author={Lyberatos, Vassilis and Kantarelis, Spyridon and Dervakos, Edmund and Stamou, Giorgos},
  journal={IEEE Access}, 
  title={Challenges and Perspectives in Interpretable Music Auto-Tagging Using Perceptual Features}, 
  year={2025},
  volume={13},
  number={},
  pages={60720-60732},
  doi={10.1109/ACCESS.2025.3555741}}

@misc{agostinelli2023musiclm,
  author = {Andrea Agostinelli and Timo I. Denk and Zalán Borsos and Jesse Engel and Mauro Verzetti and Antoine Caillon and Qingqing Huang and Aren Jansen and Adam Roberts and Marco Tagliasacchi and Matt Sharifi and Neil Zeghidour and Christian Frank},
  title = {MusicLM: Generating Music From Text},
  year = {2023},
  eprint = {2301.11325},
  archivePrefix = {arXiv},
  primaryClass = {cs.SD},
  url = {https://arxiv.org/abs/2301.11325}
}

@article{qwen2023,
  title={Qwen technical report},
  author={Bai, Jinze and Bai, Shuai and Chu, Yunfei and Cui, Zeyu and Dang, Kai and Deng, Xiaodong and Fan, Yang and Ge, Wenbin and Han, Yu and Huang, Fei and others},
  journal={arXiv preprint arXiv:2309.16609},
  year={2023}
}

@inproceedings{wu2023large,
  title={Large-scale contrastive language-audio pretraining with feature fusion and keyword-to-caption augmentation},
  author={Wu, Yusong and Chen, Ke and Zhang, Tianyu and Hui, Yuchen and Berg-Kirkpatrick, Taylor and Dubnov, Shlomo},
  booktitle={ICASSP 2023},
  pages={1--5},
  year={2023},
  organization={IEEE}
}

@article{alajanki2016benchmarking,
  title={Benchmarking music emotion recognition systems},
  author={Alajanki, Anna and Yang, Yi-Hsuan and Soleymani, Mohammad},
  journal={PloS one},
  volume={11},
  number={3},
  pages={e0150930},
  year={2016},
  publisher={ACM}
}

@article{hung2021emopia,
  title={EMOPIA: A multi-modal pop piano dataset for emotion recognition and emotion-based music generation},
  author={Hung, Hsiao-Tzu and Ching, Joann and Doh, Seungheon and Kim, Nabin and Nam, Juhan and Yang, Yi-Hsuan},
  journal={arXiv preprint arXiv:2108.01374},
  year={2021}
}

@article{chu2024qwen2,
  title={Qwen2-audio technical report},
  author={Chu, Yunfei and Xu, Jin and Yang, Qian and Wei, Haojie and Wei, Xipin and Guo, Zhifang and Leng, Yichong and Lv, Yuanjun and He, Jinzheng and Lin, Junyang and others},
  journal={arXiv preprint arXiv:2407.10759},
  year={2024}
}

@article{jacobsen2024assessing,
  title={Assessing aesthetic music-evoked emotions in a minute or less: A comparison of the GEMS-45 and the GEMS-9},
  author={Jacobsen, Peer-Ole and Strauss, Hannah and Vigl, Julia and Zangerle, Eva and Zentner, Marcel},
  journal={Musicae Scientiae},
  volume={29},
  number={1},
  pages={184--192},
  year={2025},
  publisher={SAGE Publications Sage UK: London, England}
}

@inproceedings{chen2016xgboost,
  title={Xgboost: A scalable tree boosting system},
  author={Chen, Tianqi and Guestrin, Carlos},
  booktitle={Proceedings of the 22nd acm sigkdd international conference on knowledge discovery and data mining},
  pages={785--794},
  year={2016}
}

@article{russell1980circumplex,
  title={A circumplex model of affect.},
  author={Russell, James A},
  journal={Journal of personality and social psychology},
  volume={39},
  number={6},
  pages={1161},
  year={1980},
  publisher={American Psychological Association}
}

@article{cesarini2024reverb,
  title={Reverb and Noise as Real-World Effects in Speech Recognition Models: A Study and a Proposal of a Feature Set},
  author={Cesarini, Valerio and Costantini, Giovanni},
  journal={Applied Sciences},
  volume={14},
  number={23},
  pages={11446},
  year={2024},
  publisher={MDPI}
}

@article{gao2022black,
  title={Black-box adversarial attacks through speech distortion for speech emotion recognition},
  author={Gao, Jinxing and Yan, Diqun and Dong, Mingyu},
  journal={EURASIP Journal on Audio, Speech, and Music Processing},
  volume={2022},
  number={1},
  pages={20},
  year={2022},
  publisher={Springer}
}

@article{panda2020audio,
  title={Audio features for music emotion recognition: a survey},
  author={Panda, Renato and Malheiro, Ricardo and Paiva, Rui Pedro},
  journal={IEEE Transactions on Affective Computing},
  volume={14},
  number={1},
  pages={68--88},
  year={2020},
  publisher={IEEE}
}

@article{mcinnes2018umap,
  title={Umap: Uniform manifold approximation and projection for dimension reduction},
  author={McInnes, Leland and Healy, John and Melville, James},
  journal={arXiv preprint arXiv:1802.03426},
  year={2018}
}

@inproceedings{elizalde2022clap,
  title={Clap learning audio concepts from natural language supervision},
  author={Elizalde, Benjamin and Deshmukh, Soham and Al Ismail, Mahmoud and Wang, Huaming},
  booktitle={ICASSP 2023-2023 IEEE International Conference on Acoustics, Speech and Signal Processing (ICASSP)},
  pages={1--5},
  year={2023},
  organization={IEEE}
}

@article{li2023mert,
  title={Mert: Acoustic music understanding model with large-scale self-supervised training},
  author={Li, Yizhi and Yuan, Ruibin and Zhang, Ge and Ma, Yinghao and Chen, Xingran and Yin, Hanzhi and Xiao, Chenghao and Lin, Chenghua and Ragni, Anton and Benetos, Emmanouil and others},
  journal={arXiv preprint arXiv:2306.00107},
  year={2023}
}

@inproceedings{albin2015objective,
  title={Objective study of the performance degradation in emotion recognition through the AMR-WB+ codec.},
  author={Albin, Aaron and Moore, Elliot},
  booktitle={INTERSPEECH},
  pages={1319--1323},
  year={2015}
}

@inproceedings{akman2025audio,
  title={Audio explanation synthesis with generative foundation models},
  author={Akman, Alican and Sun, Qiyang and Schuller, Bj{\"o}rn W},
  booktitle={ICASSP 2025},
  pages={1--5},
  year={2025},
  organization={IEEE}
}

@inproceedings{deng2025investigating,
  title={Investigating the Sensitivity of Pre-trained Audio Embeddings to Common Effects},
  author={Deng, Victor and Wang, Changhong and Richard, Gael and McFee, Brian},
  booktitle={ICASSP 2025},
  pages={1--5},
  year={2025},
  organization={IEEE}
}

@inproceedings{lyberatos2024perceptual,
  title={Perceptual musical features for interpretable audio tagging},
  author={Lyberatos, Vassilis and Kantarelis, Spyridon and Dervakos, Edmund and Stamou, Giorgos},
  booktitle={2024 IEEE International Conference on Acoustics, Speech, and Signal Processing Workshops (ICASSPW)},
  pages={878--882},
  year={2024},
  organization={IEEE}
}

@article{shah2021all,
  title={What all do audio transformer models hear? probing acoustic representations for language delivery and its structure},
  author={Shah, Jui and Singla, Yaman Kumar and Chen, Changyou and Shah, Rajiv Ratn},
  journal={arXiv preprint arXiv:2101.00387},
  year={2021}
}

@inproceedings{laurier2009exploring,
  title={Exploring relationships between audio features and emotion in music},
  author={Laurier, Cyril and Lartillot, Olivier and Eerola, Tuomas and Toiviainen, Petri},
  booktitle={Proceedings of the 7th Triennial Conference of European Society for the Cognitive Sciences of Music},
  pages={260--264},
  year={2009},
  organization={Citeseer}
}

@article{han2023effect,
  title={Effect of sound sequence on soundscape emotions},
  author={Han, Zhihui and Kang, Jian and Meng, Qi},
  journal={Applied Acoustics},
  volume={207},
  pages={109371},
  year={2023},
  publisher={Elsevier}
}

\end{document}